\documentclass[3p,twocolumn,sort&compress]{elsarticle}
\usepackage{graphicx, fancybox}
\usepackage{amsmath,amssymb,bm}
\usepackage[T1]{fontenc} % if needed
\usepackage[table]{xcolor}
\usepackage{ulem}

\newcommand{\red}[1]{\textcolor{red}{#1}}
\newcommand{\blue}[1]{\textcolor{blue}{#1}}

\begin{document}

\title{Distribution of Stress Tensor around Static Quark--Anti-Quark\\ from Yang-Mills Gradient Flow }

\author[osaka]{Ryosuke Yanagihara}
\ead{yanagihara@kern.phys.sci.osaka-u.ac.jp}

\author[riken]{Takumi Iritani}

\author[osaka,kek]{Masakiyo Kitazawa}

\author[osaka]{Masayuki Asakawa}

\author[riken_i,riken]{Tetsuo Hatsuda}

\address[osaka]{Department of Physics, Osaka University,
Toyonaka, Osaka 560-0043, Japan}

\address[riken]{RIKEN Nishina Center, RIKEN, Wako 351-0198, Japan}

\address[kek]{
  J-PARC Branch, KEK Theory Center,
  Institute of Particle and Nuclear Studies, KEK,
  203-1, Shirakata, Tokai, Ibaraki, 319-1106, Japan }

\address[riken_i]{
  RIKEN Interdisciplinary Theoretical and Mathematical Sciences Program
  (iTHEMS), RIKEN, Wako 351-0198, Japan}
  	
\begin{abstract}
The spatial distribution of the stress tensor around the
quark--anti-quark ($Q\bar{Q}$) pair in SU(3) lattice gauge  theory is studied. The
Yang-Mills gradient flow plays a crucial role to make the stress tensor well-defined 
and derivable from the numerical simulations on the lattice.
The resultant stress tensor with a decomposition into local principal axes shows, for the
first time, the detailed structure of the  flux tube  along the longitudinal and transverse directions  in a gauge invariant manner.
 The linear confining behavior of the $Q\bar{Q}$ potential at long distances
is derived directly from the integral of the local stress tensor.
\end{abstract}

\begin{keyword}
%\preprint{J-PARC-TH-117, QHP-365, RIKEN-iTHEMS-Report-18}
 Lattice gauge theory \sep Strong interaction \sep Gradient flow \sep Stress tensor
 \PACS 12.38.Gc,12.38.Aw,11.15.Ha
\end{keyword}

\maketitle

The energy-momentum tensor (EMT), ${\cal T}_{\mu \nu}(x)$,
in classical and quantum field theories is a special
quantity among other observables in the sense that it relates
the local properties and the global behaviors of the system in a gauge invariant manner.
A classic example is the Maxwell stress-tensor 
in electromagnetism:
$\sigma_{ij}^{\rm Maxwell}=-{\cal T}_{ij}^{\rm Maxwell} =  -(F_{i\mu} F^{\mu}_{\ j} - \frac{1}{4} \delta_{ij} F_{\mu\nu}^2)$
~\cite{Landau}.
It describes the
local response under external charges, and its integration on
the surface surrounding
a charge gives the Coulomb force acting on the charge. In quantum Yang-Mills (YM) theory,
the EMT is even more important than in the Abelian case, since
it provides gauge-invariant and non-perturbative information.

The purpose of this Letter is to explore novel aspects of EMT
in YM theory at zero temperature under the presence of static quark ($Q$) and anti-quark ($\bar{Q}$)
charges separated by a distance $R$.
In such a setup, the YM field strength is believed to be squeezed into a quasi-one-dimensional
flux tube~\cite{Nambu} and gives rise to the linear confining potential at large $R$
(see the reviews~\cite{Bali:2000gf,Greensite,Kondo:2014sta} and references therein).
Although the action density, the color electric field and the plaquettes 
 have been employed before to probe such a flux tube
 ~\cite{DiGiacomo:1990hc,Bali:1994de,Michael:1995pv,Green:1996be,Gliozzi:2010zv,Meyer:2010tw,Cea:2012qw,Cardoso:2013lla},
the present Letter is a first attempt to provide  gauge invariant 
EMT distribution  around the $Q\bar{Q}$ pair in three spatial dimensions.
The fundamental theoretical tool to make this
analysis possible is the YM gradient flow~\cite{Narayanan:2006rf,Luscher:2010iy,Luscher:2011bx},
which was recently put in practice to treat ${\cal T}_{\mu \nu}(x)$~\cite{Suzuki:2013gza,Suzuki:2016ytc}
and has been applied extensively to the equation of state of SU(3) YM theory
at finite temperature~\cite{Asakawa:2013laa,Kitazawa:2016dsl,Kamata:2016any,Kitazawa:2017qab}.

Before going into the details of our lattice study, let us first discuss
the general feature of ${\cal T}_{\mu \nu}(x)$ in the Euclidean spacetime
with $\mu,\nu=1,2,3,4$.
The local energy density and  the stress tensor read respectively as 
\begin{eqnarray}
 \varepsilon (x) &=& -  {\cal T}_{44}(x), \\
 \sigma_{ij} (x) &=& - {\cal T}_{ij}(x)\ (i,j=1,2,3).
\end{eqnarray}
The force per unit area ${\cal F}_i$,
which induces the momentum flow through a given
surface element with the normal vector $n_i$, is given by \cite{Landau}
\begin{eqnarray}
{\cal F}_i =  \sigma_{ij}n_j =  - {\cal T}_{ij} n_j.
\label{eq:F=Tn}
\end{eqnarray} 
Then the  local  principal axes and the corresponding eigenvalues
of the local stress can be obtained  by diagonalizing ${\cal T}_{ij}$:
\begin{align}
{\cal T}_{ij}n_j^{(k)}=\lambda_k n_i^{(k)} \quad (k=1,2,3),
\label{eq:EV}
\end{align}
where the strengths of the force per unit area
along the
principal axes are given by the absolute values of the
eigenvalues, $\lambda_k $.
The force acting on a test charge is
obtained by the surface integral ${F}_i =- \int_S {\cal T}_{ij} dS_j$, where $S$ is a surface surrounding the charge with the surface vector $S_j$ oriented  outward from $S$.

In quantum YM theory, obtaining  ${\cal T}_{\mu \nu}(x)$
non-perturbatively around static $Q\bar{Q}$ on the lattice
requires us to go through the following steps.

{\it The first step} is to start with  the YM gradient flow equation~\cite{Luscher:2010iy},
\begin{align}
\frac{dA_\mu(t,x)}{dt} = -g_0^2 \frac{\delta S_{\mathrm{YM}}(t)}{\delta A_\mu (t,x)},
\end{align} 
with the fictitious 5-th coordinate $t$.
The YM action $S_{\mathrm{YM}}(t)$
is composed of $A_{\mu}(t,x)$,
whose initial condition at $t=0$ is
the ordinary gauge field $A_\mu(x)$ in the four dimensional
Euclidean space.
The gradient flow for positive $t$ smooths  the gauge field
with the radius $ \sqrt{2t}$.
Then the renormalized EMT operator is defined as ~\cite{Suzuki:2013gza}
\begin{align}
 {\cal T}^{\rm R}_{\mu\nu}(x)&=\lim_{t\rightarrow0}{\cal T}_{\mu\nu}(t,x),\\
 {\cal T}_{\mu\nu}(t,x)&=\frac{U_{\mu\nu}(t,x)}{\alpha_U(t)}
  +\frac{\delta_{\mu\nu}}{4\alpha_E(t)}  [E(t,x)- \langle E(t,x)
 \rangle_0 ].
\end{align}
Here $E(t,x)=(1/4)G_{\mu\nu}^a(t,x)G_{\mu\nu}^a(t,x)$ and
$U_{\mu\nu}(t,x)=G_{\mu\rho}^a(t,x)G_{\nu\rho}^a(t,x)-\delta_{\mu\nu}E(t,x)$
with the field strength $G_{\mu\nu}^a(t,x)$ composed of the flowed gauge field $A_\mu (t,x)$.
The vacuum expectation value $\langle {\cal T}_{\mu\nu}(t,x) \rangle_0$ is normalized to be zero due to the subtraction of $\langle E(t,x) \rangle_0$.
We use the perturbative coefficients for $\alpha_U(t)$ and
$\alpha_E(t)$~\cite{Suzuki:2013gza} in the following analysis.
Thermodynamic quantities in SU(3) YM  theory 
have been shown to be accurately obtained with this EMT operator
with smaller statistics
than with the previous methods~\cite{Asakawa:2013laa,Kamata:2016any,Kitazawa:2016dsl}.

{\it The second step} is to prepare a static $Q\bar{Q}$ system on the lattice.
We use the rectangular  Wilson loop $W(R,T)$
with static  color charges at
 $\vec{R}_{\pm}=(0,0, \pm  R/2)$ and in the temporal interval $[-T/2, T/2]$.
Then the  expectation value of ${\cal T}_{\mu\nu}(t,x)$ around the $Q\bar{Q}$ 
is obtained by \cite{Luscher:1981}
\begin{align}
  \langle {\cal T}_{\mu\nu}(t,x)\rangle_{Q\bar{Q}} =
  \lim_{T\to\infty} \frac{\langle  {\cal T}_{\mu\nu}(t,x) W(R,T)\rangle_0}
      {\langle W(R,T)\rangle_0} ,
      \label{eq:<T>_W} 
\end{align}
where $T \rightarrow \infty$ is to pick up the ground state of
$Q\bar{Q}$.
The measurements of ${\cal T}_{\mu\nu}(t,x)$ for different values of $t$
are made at the mid temporal plane $x_{\mu} =(\vec{x}, x_4=0)$,
while $W(R,T)$ is defined at $t=0$.

{\it The final step} is to obtain the renormalized  EMT distribution  around
${Q\bar{Q}} $ from the lattice data by taking double limit~\cite{Asakawa:2013laa,Kitazawa:2016dsl},
 \begin{eqnarray}
 \langle {\cal T}^{\rm R}_{\mu\nu}(x)\rangle_{Q\bar{Q}}
  = \lim_{t \rightarrow 0}  \lim_{a \rightarrow 0}
  \langle {\cal T}_{\mu\nu}(t,x)\rangle_{Q\bar{Q}}^{\rm lat}.
\label{eq:limits}
\end{eqnarray}
In lattice simulations we measure
$\langle {\cal T}_{\mu\nu}(t,x)\rangle_{Q\bar{Q}}^{\rm lat}$ at finite $t$ and $a$, and 
make an extrapolation to $(t,a)=(0,0)$ according to the formula~\cite{Kitazawa:2016dsl,Kitazawa:2017qab},
\begin{align}
\langle {\cal T}_{\mu\nu}(t,x)\rangle_{Q\bar{Q}}^{\rm lat}
\simeq  \langle {\cal T}^{\rm R}_{\mu\nu}(x)\rangle_{Q\bar{Q}} + b_{\mu
\nu}(t){a^2} + c_{\mu \nu} t,
\label{eq:ansatz}
\end{align}
where $b_{\mu \nu}(t)$ and $c_{\mu \nu} $ are contributions from
lattice discretization effects and the dimension six
operators, respectively.

\begin{table*}
 \centering
 \caption{
   First five rows are the simulation parameters on the lattice.
   Spatial size of the Wilson loop $R$ 
     is shown in the lattice and physical units.
   Temporal size of the Wilson loop is set to $T$
     unless otherwise stated.
 }
\begin{tabular}{|cccr|ccc|c|}
 \hline
 $\beta$ & $a~[\mathrm{fm}]$ & $N_\mathrm{size}^4$ & $N_{\rm conf}$
 &    & $R/{a}$  &  & $T/{a}$     \\
 \hline
 6.304 & 0.058 & 48$^4$ & 140   & 8  & 12 & 16 & 8    \\
 6.465 & 0.046 & 48$^4$ & 440    & 10 & -- & 20 & 10 \\
 6.513 & 0.043 & 48$^4$ & 600    & -- & 16 & -- & 10 \\
 6.600 & 0.038 & 48$^4$ & 1,500  & 12 & 18 & 24 & 12 \\
 6.819 & 0.029 & 64$^4$ & 1,000 & 16 & 24 & 32 & 16   \\
 \hline

   & & \multicolumn{2}{c|}{$R ~[\mathrm{fm}]$}  &   0.46 & 0.69  & 0.92 &    \\
 \hline
 \end{tabular}
 \label{table:param}
\end{table*}

The numerical simulations in SU(3) YM theory are performed on the
four-dimensional Euclidean lattice with the Wilson gauge action
and the periodic boundary condition.
Shown in Table~\ref{table:param} are five different inverse couplings
$\beta=6/g_0^2$ and  corresponding  lattice spacings $a$
determined from the $w_0$-scale~\cite{Borsanyi:2012zs,Kitazawa:2016dsl}.
The lattice size $N_{\rm size}^4$, and the
number of gauge  configurations $N_{\rm conf}$ are also summarized in the table.
Gauge configurations are generated by the same procedure as 
in~\cite{Kitazawa:2016dsl} with the  separation of
$200$ ($100$) sweeps on the $64^4$ ($48^4$) lattice.
\footnote{Our simulation on fine lattices may suffer from 
	the topological freezing ~\cite{Endres:2016rzj}.
	However, from the analysis with the gauge configurations
	used for the scale setting in ~\cite{Kitazawa:2016dsl},
	we have checked that the dependence
	of $E(t, x)$ on the topological sector is less than 
	1\% on $64^4$ lattice at $\beta = 6.88$. 
       	With a reasonable assumption that  the dependence of Eq.~(\ref{eq:<T>_W}) on different  
	topological sectors is of  the same order, 
	the topological freezing is likely to be less than  the statistical errors in the present study.} 
	Statistical errors are estimated by the jackknife method with $20$
	jackknife bins at which the errors saturate. In the flow equation,
	the Wilson gauge action is used for $S_{\mathrm{YM}}(t)$, while
	the clover-type representation is adopted for $G^a_{\mu\nu}(t,x)$
	in ${\cal T}_{\mu\nu}(t,x)$.

Other than the gradient flow for ${\cal T}_{\mu\nu}(t,x)$ described above,
we adopt the standard APE smearing for each spatial link along the Wilson loop 
~\cite{Albanese:1987ds} with the same smearing parameter
as in ~\cite{Takahashi:2002bw}
to enhance the coupling of $W(R,T)$ to the $Q\bar{Q}$ ground state.
We keep $a \sqrt{N_{\rm APE}}$ which is  proportional to the transverse size of the spatial links ~\cite{Narayanan:2006rf,Takahashi:2002bw} to be approximately constant  
 by changing the  iteration number $N_{\rm APE}$.
Also, to reduce the statistical noise, we adopt the
standard multi-hit procedure by replacing each temporal link 
by its mean-field value~\cite{Parisi:1983hm,Bali:1994de}.

\begin{figure}[t]
  \centering
  \includegraphics[width=0.47\textwidth,clip]{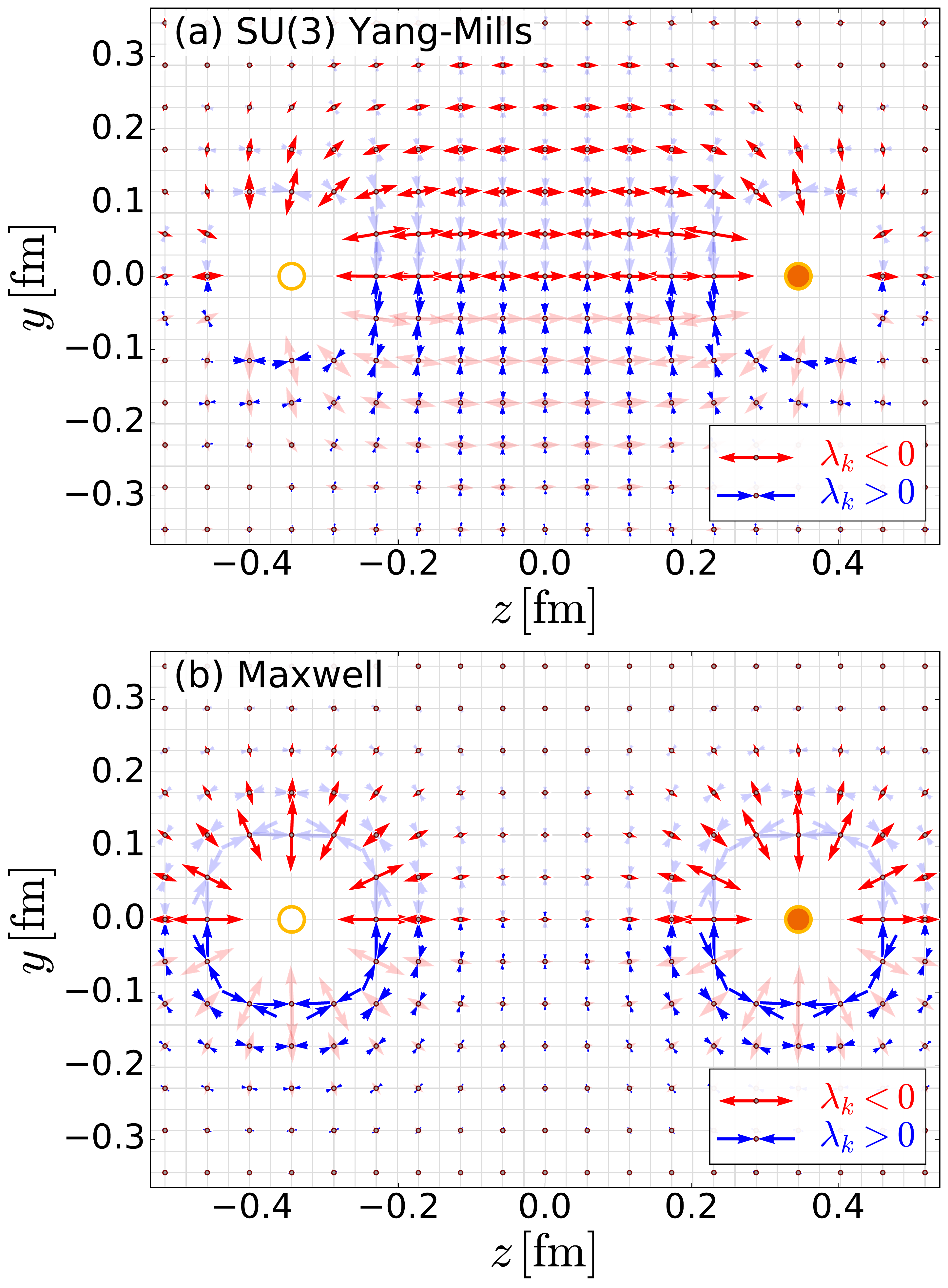}
    \caption{(a) Distribution of the principal axes of ${\cal T}_{ij}$ for a
     $Q\bar{Q}$ system separated by $R=0.69$ fm in SU(3) Yang-Mills theory.
      The lattice spacing and the flow time are fixed to be
      $a=0.029$~fm and $t/a^2=2.0$, respectively.
   (b) Distribution of the principal axes of the ${\cal T}_{ij}$
 in classical electrodynamics between opposite charges.
 In both figures,  the red (blue) arrows  in  the upper (lower) half plane are 
 highlighted. }
 \label{fig:stress-distribution}
\end{figure}

\begin{figure*}[t]
 \centering
 \includegraphics[width=0.99\textwidth,clip]{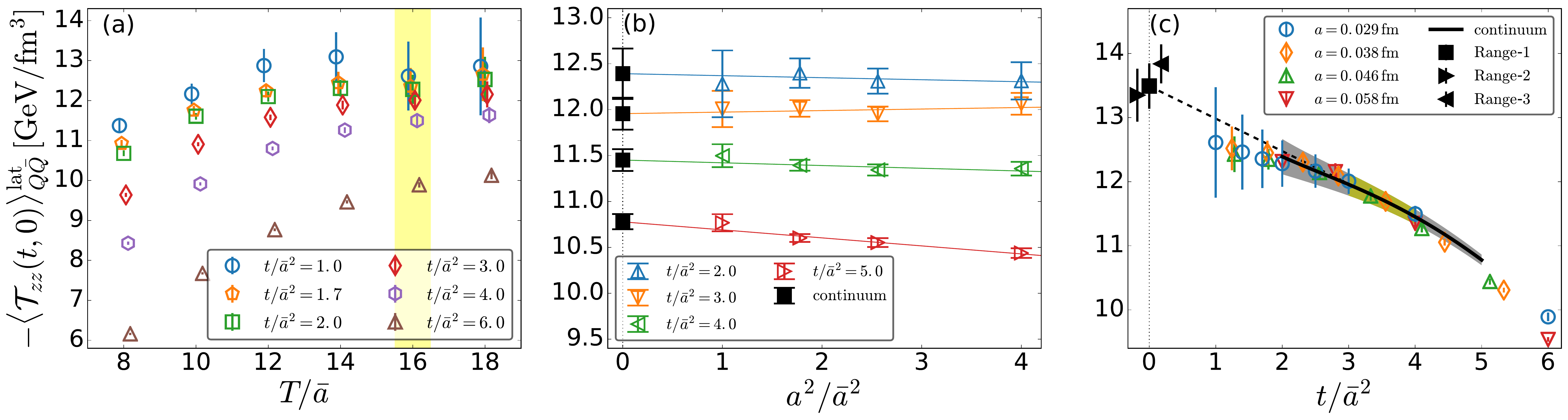}
 \caption{
 $zz$ component of the  stress tensor at $\vec{x}=0$,
 for various $a$ and $t$ with  $R=0.46$~fm.
 (a) Open symbols  with the statistical errors are
 $\langle {\cal T}_{zz}(t,0)\rangle^{\rm lat}_{Q\bar{Q}}$
 as a function of $T/\bar{a}$  with $a= \bar{a}=0.029\ {\rm fm} $.
 To take the double limit, the data at
 $T/\bar{a}=16$ indicated by the yellow band is used.
 (b) Open symbols with errors denote  $\langle {\cal T}_{zz}(t,0)\rangle^{\rm lat}_{Q\bar{Q}}$
 as a function of  $a^2/\bar{a}^2$.  The filled black symbols are
 the  results of the $a\to 0$ limit for each $t$.
 (c) Open symbols with errors are    $\langle {\cal T}_{zz}(t,0)\rangle^{\rm lat}_{Q\bar{Q}}$ as a function of
 $t/\bar{a}^2$   for different $a$. The solid line corresponds to the
 result of $a\to 0$ limit in the interval $2 \le t/\bar{a}^2 \le 5$ with shaded band being statistical
 error. The filled black symbols are the results of the $t\to 0$
 extrapolation from three different ranges of $t$.}
 \label{fig:mix}
\end{figure*}

We consider three $Q\bar{Q}$ distances ($R=0.46, 0.69, 0.92$ fm),
which are comparable to the typical scale of strong interaction.
These values as well as  the corresponding dimensionless distances $R/a$
are  summarized in Table~\ref{table:param}.
While the largest $R$ is half the spatial lattice extent $aN_{\rm size}$
for the two finest lattice spacings, effects of the periodic boundary are
known to be well suppressed even with this setting~\cite{Bali:1994de}.
A measure of the ground state saturation in  the $Q\bar{Q}$ system reads
\begin{align}
{P}(R, T) = {C_0 (R)  e^{-V(R)  T} }/{\langle W(R,T)\rangle_0}
\end{align}
with the ground-state potential $V(R)$ and the ground-state overlap $C_0 (R)$
 obtained at large  $T$ ~\cite{Bali:1994de}.
Using the data at $a=0.038$ fm with $N_{\rm APE} =160$,
we found $|1-{P} (R, T)| < {0.5} \%$ as long as   $T > 0.19$ fm 
for  all $R$ in Table~\ref{table:param}.  
 By keeping $T\simeq0.46$~fm to extract observables 
    as  shown in the last column of Table~\ref{table:param}, the ground state saturation of the 
     Wilson loop is, therefore, secured in our 
     simulations.\footnote{An alternative estimate of the ground state saturation 
     is obtained through the excitation energy of a bosonic string:   $\Delta E_n = \pi n / R \ (n=,1,2,\cdots)$
         \cite{Luscher:2002qv}.  
     By taking $n=2$, which corresponds to the  excitation with the same symmetry ($\Sigma_g^+$) as the ground state  \cite{Juge:2002br},
     the excited state is suppressed at least by a factor  $ \exp(-(2\pi/R) T) = \exp( -\pi ) \simeq 4\%$ for $T=0.46$ fm and 
     $R=0.92$ fm, even if $C_{n=2} (R) \sim C_{n=0} (R)$.}

To avoid the artifact due to finite $a$ and
the over-smearing of the gradient flow~\cite{Kitazawa:2016dsl,Kitazawa:2017qab},
we need to choose an  appropriate window of $t$ satisfying the 
 condition  ${a}/{2} \lesssim  \rho  \lesssim L$.
 Here $\rho \equiv \sqrt{{2t}}$ is the flow radius, and 
$L\equiv {\rm min} (|\vec{x}-\vec{R}_{+}|, |\vec{x}-\vec{R}_{-}|, T/2)$
is the minimal distance between $x_{\mu}=(\vec{x},0)$ and the Wilson loop.

Before taking the double limit in Eq.~(\ref{eq:limits}), we 
illustrate a qualitative feature of
 the distribution of ${\cal T}_{ij}$ around  $Q\bar{Q}$ in YM theory
at fixed $a$ and $t$ by considering the case $a =0.029\ {\rm fm}$
with $R=0.69\ {\rm fm}$, $T=0.46\ {\rm fm}$, and $t/{a}^2 = 2.0$.
Shown in Fig.~\ref{fig:stress-distribution}  are
 the two eigenvectors in Eq.~(\ref{eq:EV}) along with the principal axes of the local stress. 
 The other eigenvector is perpendicular to the   $y$-$z$ plane.
 The eigenvector with negative (positive) eigenvalue 
   is   denoted by the red outward (blue inward) arrow with its length proportional to $\sqrt{|\lambda_k |}$:
\begin{eqnarray}
 \red{\leftarrow} \hspace{-0.05cm} {\tiny{\circ}} \hspace{-0.05cm}  \red{\rightarrow} : \lambda_k<0, 
 \qquad  \blue{\rightarrow} \hspace{-0.05cm}  {\tiny{\circ}}  \hspace{-0.05cm}  \blue{\leftarrow}: \lambda_k>0.
\end{eqnarray}
Neighbouring volume elements  are pulling (pushing) with each other along the direction of
   red (blue) arrow according to Eq.~(\ref{eq:F=Tn}).
   
The spatial regions near $Q$ and $\bar{Q}$, which 
would suffer from over-smearing, are excluded in the figure.
Spatial structure of the flux tube is clearly
revealed through the 
stress tensor in Fig.~\ref{fig:stress-distribution}~(a) in a gauge 
invariant way.
This is in contrast to the same plot of the principal axes of ${\cal T}_{ij}$
for opposite charges in classical electrodynamics given in Fig.~\ref{fig:stress-distribution}~(b).

Let us now  turn to  the mid-plane between the $Q\bar{Q}$ pair
and extract the stress-tensor distribution  at $\vec{x}=(x,y,0)$
by taking the double limit Eq.~(\ref{eq:limits}).
In Fig.~\ref{fig:mix}~(a), we show an example of the $T$ dependence of
$\langle {\cal T}_{zz}(t,0)\rangle_{Q\bar{Q}}^{\rm lat}$, which gives the largest eigenvalue
on the mid-plane,  for  several values of $t$ with $a=0.029\ {\rm fm}\ (\equiv \bar{a})$.
The figure indicates that
significant $T$-dependence arises owing to over-smearing of the gradient flow
for  $t/\bar{a}^2 \sim 6$  (i.e., $\rho \sim \sqrt{12}\bar{a}=0.10$ fm)
already around $T =16 \bar{a}=0.46$ fm.
This is so  for the same $t/\bar{a}^2$   in all other cases in Table~\ref{table:param}.
On the other hand, under-smearing of the gradient flow on the coarsest lattice
becomes significant for $t/\bar{a}^2=2$ ($\rho=a=0.058$ fm).
Therefore, in the following analysis, we focus on the data in
the interval $2 \leq t/\bar{a}^2 \leq 5$
($2 \bar{a}  \leq \rho\leq \sqrt{10} \bar{a}$), which satisfies
${a}/{2} \lesssim  \rho  \lesssim L$ with margin.

In Fig.~\ref{fig:mix}~(b), we show $\langle {\cal T}_{zz}(t,0)\rangle_{Q\bar{Q}}^{\rm lat}$
as a function of  the dimensionless ratio  $a^2/\bar{a}^2$  by the open triangles
for different  values of $t$.  The continuum limit ($a\rightarrow 0$)  with fixed $t$
is taken by using these data together with the formula Eq.~(\ref{eq:ansatz}). The results are
shown by the filled black squares with error bars.

\begin{figure*}[t]
  \centering
  \includegraphics[width=0.95\textwidth,clip]{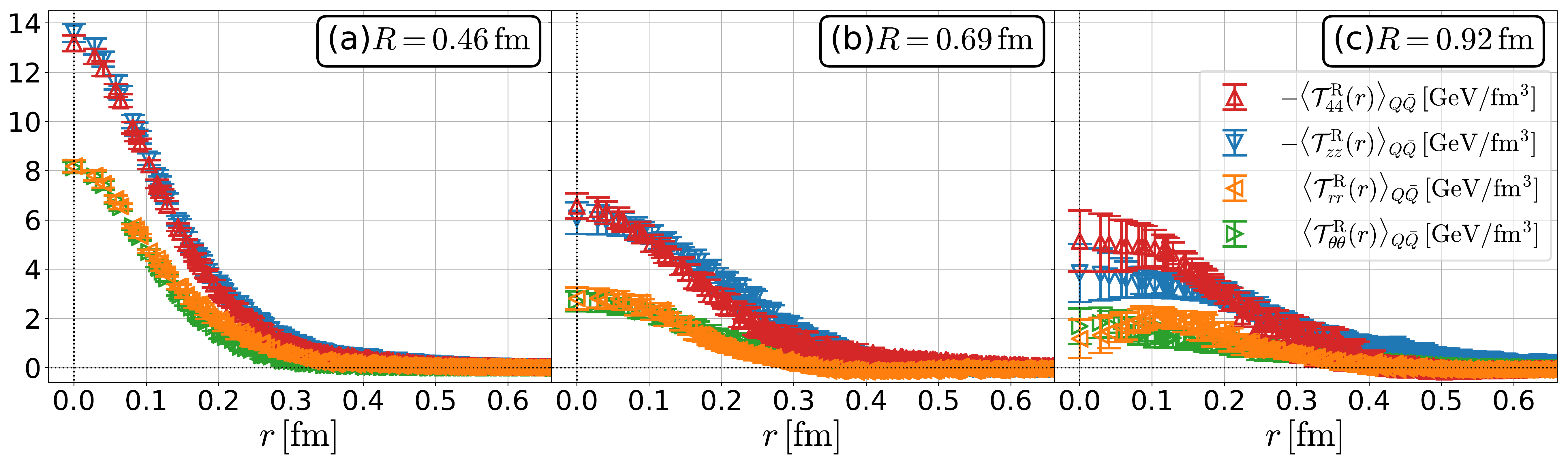}
  \caption{
EMT distribution on the mid-plane after the double limit
 $-\langle{\cal T}^{\rm R} _{cc}(r) \rangle_{Q\bar{Q}} $
 and $-\langle{\cal T}^{\rm R} _{44}(r) \rangle_{Q\bar{Q}} $
 in the cylindrical coordinate system for three different values of the $Q\bar{Q}$ distance $R$.
  \label{fig:mid}
  }
\end{figure*}

In Fig.~\ref{fig:mix}~(c), the open symbols with error bars correspond
to the original data for various values of  $a$ and $t/\bar{a}^2$.
The result of the continuum limit  in the interval $2 \le t/\bar{a}^2 \le 5$
is denoted  by the black solid line with the shaded error band.
The $t\rightarrow 0$ limit  is carried out by using the values in the continuum limit
according to Eq.~(\ref{eq:ansatz}) with $a=0$.
As a  most conservative range for the extrapolation, we take  $3 \leq
t/\bar{a}^2 \leq 4$ (Range-1). Also, to estimate the systematic errors
from the extrapolation, we consider two different ranges by changing the upper and lower limits:
$2 \leq t/\bar{a}^2 \leq 4$ (Range-2) and
$3 \leq t/\bar{a}^2 \leq 5$ (Range-3).
The resulting values of $\langle {\cal T}_{zz}^{\rm R}(0)\rangle_{Q\bar{Q}}$  after the double limit
are shown by the filled black symbols at $t=0$. The dashed line
corresponds to the extrapolation with Range-1.

\begin{figure}
  \centering
  \includegraphics[width=0.47\textwidth,clip]{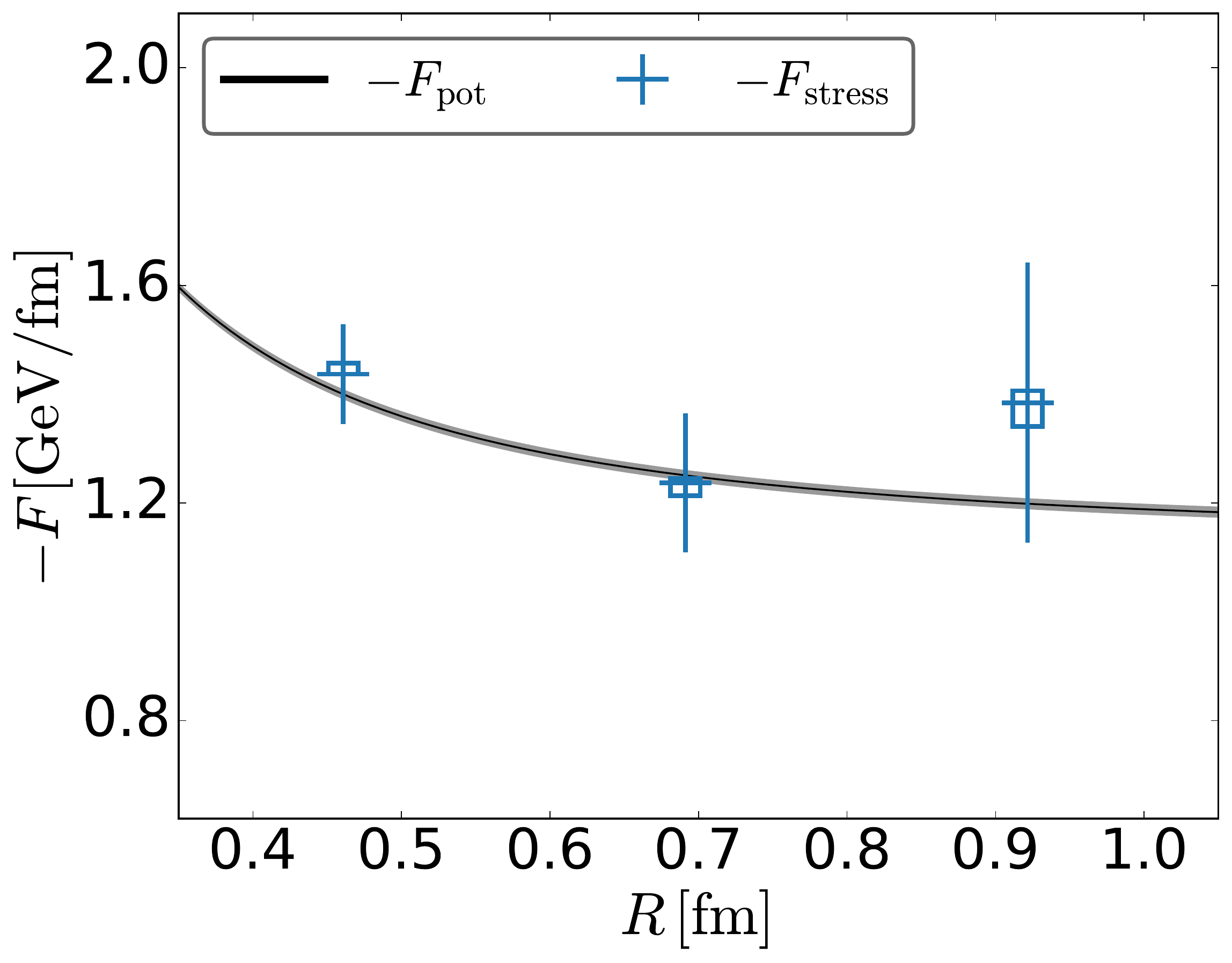}
  \caption{
  $R$ dependence of the $Q\bar{Q}$ forces,
   $-F_{\rm stress}$ and $-F_{\rm pot}$, obtained by the Wilson loop
 and the stress tensor, respectively.
 Error bars and rectangular boxes for the latter represent the  statistical and systematic errors,
 respectively.
 \label{fig:R}
  }
\end{figure}

The double limit  for general $\vec{x}=(x,y,0)$ on the mid-plane can be
carried out essentially through the same procedure with a few extra steps.
First of all, the cylindrical coordinate system $c=(r,\theta, z)$  with
$r=\sqrt{x^2+y^2}$ and $0 \le \theta < 2 \pi$ is useful for the present
$Q\bar{Q}$ system.
On the mid-plane we have
    $\langle {\cal T}_{c c'}(t,x) \rangle_{Q\bar{Q}}^{\rm lat} =
    {\rm diag} ( \langle{\cal T}_{rr} (t,r)\rangle_{Q\bar{Q}}^{\rm lat},
    \langle {\cal T}_{\theta\theta} (t,r)\rangle_{Q\bar{Q}}^{\rm lat},
    \langle {\cal T}_{zz} (t,r)\rangle_{Q\bar{Q}}^{\rm lat})$.
Next, we need data at the same $r$ for different $a$
to take the continuum limit.
We consider the values of $r$ at which the lattice data are
  available on the finest lattice. To obtain the data 
  at these $r$ on lattices with different $a$, we interpolate
the
lattice data $\langle {\cal T}_{c c}(t,r) \rangle_{Q\bar{Q}}^{\rm lat}$
 and $\langle {\cal T}_{44}(t,r) \rangle_{Q\bar{Q}}^{\rm lat}$
with the commonly used functions
to parametrize the transverse profile of the flux tube:
$f_{\rm Bessel} (r)= A_0 K_0\big(\sqrt{Br^2+C}\big)$
with the 0th-order modified Bessel function $K_0(x)$~\cite{Clem:1975}
and $f_{\rm exp} (r)= (A_0+A_1r^2) e^{(-2\sqrt{r^2+B^2}+2B)/C}$~\cite{Cardoso:2013lla}.
Once it is done, the $t \to 0$ limit is taken in the same way as
explained above.

In Fig.~\ref{fig:mid}, we show the $r$ dependence of the resulting EMT
(the stress tensor  $ - \langle {\cal T}^{\rm R}_{cc}(r) \rangle_{Q\bar{Q}} $ and the
energy density  $-\langle {\cal T}^{\rm R}_{44}(r)\rangle_{Q\bar{Q}} $).
From the figure, one finds several noticeable features:
\begin{itemize}
\item[(i)] Approximate degeneracy between  temporal and longitudinal components 
is found for a wide range of $r$:  
$ \langle {\cal T}^{\rm R}_{44}(r)  \rangle_{Q\bar{Q}}  \simeq
\langle {\cal T}^{\rm R}_{zz}(r) \rangle_{Q\bar{Q}} < 0 $.  This feature is compatible with the
 leading-order prediction of the worldsheet theory of QCD string~\cite{Meyer:2010tw}.
We also find 
$ \langle {\cal T}^{\rm R}_{rr}(r)  \rangle_{Q\bar{Q}}  \simeq
\langle{\cal T}^{\rm R}_{\theta \theta}(r)  \rangle_{Q\bar{Q}} >0  $, which 
does not have simple interpretation except at $r=0$.
%These degeneracy and sepatation are non-trivial consequences in non-Abelian gauge theory.

\item[(ii)] The scale symmetry broken in the YM vacuum (the trace anomaly) is
partially restored inside the flux tube, which arises in the numerical results,
$ \langle {\cal T}^{\rm R}_{\mu \mu}(r) \rangle_{Q\bar{Q}} =
\langle {\cal T}^{\rm R}_{44}(r) +  {\cal T}^{\rm R}_{zz}(r) +
{\cal T}^{\rm R}_{rr}(r)  +  {\cal T}^{\rm R}_{\theta \theta}(r) \rangle_{Q\bar{Q}} <  0 $.
This is in sharp  contrast to  the case of classical electrodynamics;
 ${\cal T}_{44}(r)={\cal T}_{zz}(r)=-{\cal T}_{rr}(r)=
-{\cal T}_{\theta \theta}(r)$ and    ${\cal T}_{\mu \mu}(r)=0$ for all $r$.
\item[(iii)] Each component of EMT at $r=0$ decreases as
  $R$ becomes larger, while the transverse radius of the flux tube,
  typically about $0.2$~fm, seems to increase for large $R$~\cite{Cardoso:2013lla,Luscher:1981,Gliozzi:2010zv},
  although the statistics is not enough to discuss
  the radius quantitatively.
 \end{itemize} 

Finally, we consider a non-trivial relation between the
force acting on the charge located at $z>0$
evaluated by the $Q\bar{Q}$ potential through
\begin{align}
F_{\rm pot}=-dV(R)/dR
\end{align}
and the force 
evaluated by the surface integral of the stress-tensor 
surrounding the charge,
\begin{align}
F_{\rm stress} = - \int  \langle {\cal T}_{zj}(x) \rangle_{Q\bar{Q}}\  dS_j .
\end{align}
For $F_{\rm pot}$,
we fit the numerical data of  $V(R)$ obtained  from the Wilson loop
at $a=0.038$ fm by the Cornell parametrization, $V_{\rm
Cornell}(R)=-{\cal A}/R+\sigma R + {\cal B}$.
Note that $V(R)$ at this lattice spacing
is shown to be already close to the continuum limit~\cite{Bali:2000gf}.
For  $F_{\rm stress}$, we take the mid-plane for  the surface integral:
$F_{\rm stress} = 2 \pi \int_0^{\infty} \langle {\cal T}_{zz}(r)
\rangle_{Q\bar{Q}}  \ r dr $.
Here $\langle {\cal T}_{zz}(r) \rangle_{Q\bar{Q}}$ is obtained by fitting Fig.~\ref{fig:mid}
with either $f_{\rm Bessel}(r)$ or  $f_{\rm exp}(r)$.
In Fig.~\ref{fig:R}, $- F_{\rm pot}$ and $- F_{\rm stress}$
thus obtained are shown by the solid line and the
horizontal bars, respectively.
$-F_{\rm pot}$ increases as ${\cal A}/R^2$ at short distance and approaches
the string tension $\sigma$ at large distance.
For $-F_{\rm stress}$,  we take into account not only the statistical error but
also the systematic errors from the double limit and the fitting in terms of $f_{\rm Bessel,exp}(r)$.
The agreement between the two quantities within the errors
is a first numerical evidence that the ``action-at-a-distance''
$Q\bar{Q}$ force can be described
by the  local properties of the stress tensor in YM theory.

In this Letter, we have performed a first study on the spatial distribution of EMT
around the  $Q\bar{Q}$ system in SU(3) lattice gauge theory. The
EMT operator defined through the YM gradient flow plays a crucial role here.
The transverse structure of the stress-tensor distribution in the mid-plane is analyzed in detail
by taking the continuum limit and the zero flow-time limit successively.
The linear confining behavior of the $Q\bar{Q}$ potential at long distances
can be shown to be reproduced by the surface integral of the stress tensor.
Further details of the stress-tensor  distribution, not only in the
transverse direction but also in the longitudinal direction,
and $R$ dependence of the transverse radius~\cite{Cardoso:2013lla,Luscher:1981,Gliozzi:2010zv},
will
be reported in a forthcoming publication~\cite{Yanagihara:prep}.
There are also  interesting future problems to be studied on the basis
of the formalism presented in this Letter: Those include
the applications to the $QQQ$ system~\cite{Takahashi:2002bw} and to the $Q\bar{Q}$ system
at finite temperature~\cite{Cea:2015wjd} as well as the generalization
to full QCD~\cite{Cea:2017ocq} with the QCD flow equation~\cite{Makino:2014taa,Taniguchi:2016ofw}.

The authors thank K.~Kondo, F.~Negro, A.~Shibata, and H.~Suzuki for discussions.
Numerical simulation was carried out on IBM System Blue Gene Solution
at KEK under its Large-Scale Simulation Program (No.~16/17-07).
This work was supported by JSPS Grant-in-Aid for Scientific Researches,
  17K05442, 25287066 and 18H05236. T.H. is grateful to the Aspen Center for Physics, 
supported in part by NSF Grants PHY1607611.

\end{document}